\documentclass[twocolumn]{aastex6}

\shorttitle{The Impact of Stellar Distances on Habitable Zone Planets}
\shortauthors{Stephen R. Kane}

\begin{document}

\title{The Impact of Stellar Distances on Habitable Zone Planets}

\author{Stephen R. Kane}
\affil{Department of Earth Sciences, University of California,
  Riverside, CA 92521, USA}
\email{skane@ucr.edu}


\begin{abstract}

Among the most highly valued of exoplanetary discoveries are those of
terrestrial planets found to reside within the Habitable Zone (HZ) of
the host star. In particular, those HZ planets with relatively bright
host stars will serve as priority targets for characterization
observations, such as those involving mass determinations,
transmission spectroscopy, and direct imaging. The properties of the
star are greatly affected by the distance measurement to the star, and
subsequent changes to the luminosity result in revisions to the extent
of the HZ and the properties of the planet. This is particularly
relevant in the realm of {\it Gaia} which has released updated stellar
parallaxes for the known exoplanet host stars. Here we provide a
generalized formulation of the effect of distance on planetary system
properties, including the HZ. We apply this methodology to three known
systems and show that the recent {\it Gaia} Data Release 2 distances
have a modest effect for TRAPPIST-1 but a relatively severe effect for
Kepler-186 and LHS~1140.

\end{abstract}



\section{Introduction}
\label{intro}

At the present time, the vast majority of exoplanets have been
detected via indirect methods, most particularly through the transit
and radial velocity (RV) methods. The derived properties of the planet
and the surrounding environment are thus highly dependent upon the
ability to accurately constrain the host star parameters. As such,
there has been a concerted effort to maximize the precision of stellar
properties \citep[e.g.,][and references therein]{hub14,boy15,san17}.
A major impediment to constructing a self-consistent stellar model for
a given star is poor distance estimates since the calculated intrinsic
luminosity of the star is tied to distance and flux
measurements. Fortunately, missions such as {\it Hipparcos}
\citep{van07} and {\it Gaia} \citep{pru16} have dramatically improved
our knowledge of stellar distances and provided subsequent
improvements to stellar property estimates. In particular, the
distances from {\it Gaia} Data Release 2 (DR2) \citep{bro18} have
already had an enormous impact on estimates of stellar properties
\citep{van18}, such as those derived for the {\it Kepler} host stars
by \citet{ber18}.

As well as alterations to the fundamental properties of the star and
planet, a change in distance to a planetary system also alters the
system flux environment since that depends on the measured stellar
luminosity. In that regard, the greatest impact is on the boundaries
of the Habitable Zone (HZ) and can determine if planets lie inside or
outside of that region. The HZ is described in detail by
\citet{kas93,kop13,kas14,kop14}, is graphically illustrated by the
Habitable Zone Gallery \citep{kan12a}, and was utilized to create a
catalog of {\it Kepler} candidates that reside in the HZ
\citep{kan16}. In these previous works, the boundaries of the HZ are
divided into two major categories. The ``conservative'' HZ (CHZ) uses
the runaway greenhouse and maximum CO$_2$ greenhouse criteria to
delimit the boundaries of the region. The ``optimistic'' HZ (OHZ)
extends the width of the HZ by accounting for best-case scenarios
whereby Venus and Mars could have retained liquid surface water. The
circumstellar CHZ and OHZ boundaries are primarily determined as a
function of the luminosity and effective temperature of the host star,
the accuracy of which is paramount for assessing the number of HZ
planets and calculations of $\eta$-Earth \citep{kan14}.

In this paper, we provide a methodology for determining the impact of
a change in a star's measured distance on the planetary system
properties including those of the star, planet, and the HZ. The
methodology is derived and quantified in Section~\ref{distance}
together with a demonstration of relative changes in the various
system parameters. These relationships are applied to several case
studies in Section~\ref{casestudies}, including the TRAPPIST-1,
Kepler-186, and LHS~1140 planetary systems. We discuss additional
caveats related to changes in distance measurements that need to be
considered in Section~\ref{discussion} and conclude in
Section~\ref{conclusions} with implications of the methodology for
other systems.


\section{How New Distances Change System Parameters}
\label{distance}

We start with the impact of stellar distance on the stellar
luminosity. The relationship between stellar distance and incident
flux, $F$, is given by
\begin{equation}
  F = \frac{L_\star}{4 \pi d^2_\star}
  \label{flux}
\end{equation}
where $L_\star$ is the stellar luminosity and $d_\star$ is the
distance from the observer to the star. For a given flux received at
Earth, the fractional change in luminosity caused by a revised
distance of $d_\star^\prime$ is given by
\begin{equation}
  \frac{\Delta L_\star}{L_\star} = \frac{d_\star^{\prime 2} -
    d_\star^2}{d_\star^2} = \left( \frac{d_\star^\prime}{d_\star}
  \right)^2 - 1
  \label{luminosity}
\end{equation}

The luminosity of the star is determined via the Stefan-Boltzmann law
applied to stars
\begin{equation}
  L_\star = 4 \pi R_\star^2 \sigma T_\mathrm{eff}^4
\end{equation}
where $R_\star$ is the stellar radius and $T_\mathrm{eff}$ is its
effective temperature. Assuming a $T_\mathrm{eff}$ that is unaffected
by stellar distance (such as extraction from spectral analysis), then
the radius is given by
\begin{equation}
  R_\star = \sqrt{ \frac{L_\star}{4 \pi \sigma T_\mathrm{eff}^4} }
\end{equation}
Based on a revised luminosity from Equation~\ref{luminosity},
$L_\star^\prime$, the fractional change in stellar radius is given by
\begin{equation}
  \frac{\Delta R_\star}{R_\star} = \frac{\sqrt{L_\star^{\prime}} -
    \sqrt{L_\star}}{\sqrt{L_\star}} = \sqrt{
    \frac{L_\star^{\prime}}{L_\star} } - 1
\end{equation}
The consequence of this change in stellar radius on planetary radius,
$R_p$, is determined from its relationship to transit depth
\begin{equation}
  \Delta F = \left( \frac{R_p}{R_\star} \right)^2
\end{equation}
which is a measurement that does not depend on stellar distance. The
fractional change in planet radius is then
\begin{equation}
  \frac{\Delta R_p}{R_p} = \frac{\Delta R_\star}{R_\star}
\end{equation}
A more dramatic change occurs with the planetary bulk density since
$\rho_p \propto R_p^{-3}$, leading to a fractional change in density
of
\begin{equation}
  \frac{\Delta \rho_p}{\rho_p} = \frac{R_p^{\prime -3} -
    R_p^{-3}}{R_p^{-3}} = \left( \frac{R_p^\prime}{R_p}
  \right)^{-3} - 1
  \label{density}
\end{equation}
where $R_p^\prime$ is the revised planetary radius.

As seen in Equation~\ref{flux}, the flux is linearly proportional to
the stellar luminosity. Therefore, the fractional change in flux
received by the planet is
\begin{equation}
  \frac{\Delta F_p}{F_p} = \frac{\Delta L_\star}{L_\star}
\end{equation}
The effect of the change in stellar parameters on the HZ is derived
using the polynomial relations described in
\citet{kop13,kop14}. Specifically, the HZ distance, $d$, is
specified by
\begin{equation}
  d_\mathrm{HZ} = \left( \frac{L_\star/L_\odot}{S_\mathrm{eff}}
  \right)^{0.5} \ \mathrm{AU}
  \label{hzeq}
\end{equation}
where the stellar flux, $S_\mathrm{eff}$, is given by
\begin{equation}
  S_\mathrm{eff} = S_{\mathrm{eff} \odot} + c_1 T_\star + c_2
  T_\star^2 + c_3 T_\star^3 + c_4 T_\star^4
\end{equation}
and $T_\star = T_\mathrm{eff} - 5780$~K. The polynomial coefficients
for the CHZ and OHZ can be found in \citet{kop14}. Based on
Equation~\ref{hzeq}, the HZ distances have an identical
proportionality, as with $R_\star$, to $\sqrt{L_\star}$. The
fractional change in the HZ boundaries is then
\begin{equation}
  \frac{\Delta d}{d} = \sqrt{ \frac{L_\star^{\prime}}{L_\star} } - 1 =
  \frac{\Delta R_\star}{R_\star}
\end{equation}
This relationship applies to both the CHZ and OHZ boundaries.

\begin{figure}
  \includegraphics[angle=270,width=8.2cm]{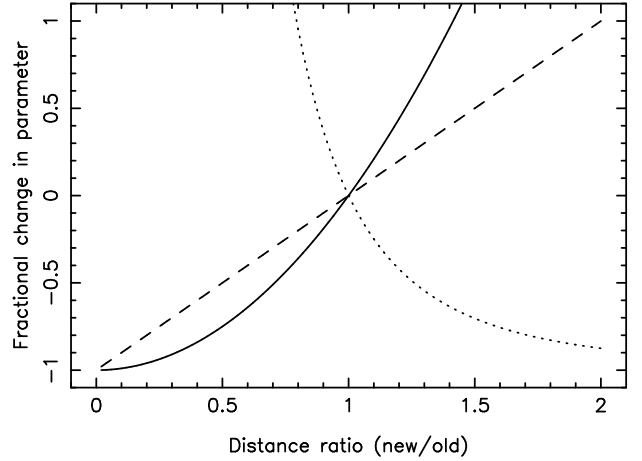}
  \caption{The fractional change in system parameters as a function of
    the ratio of the new to the old stellar distance. The solid curves
    applies to luminosity, $L_\star$, and flux received by the planet,
    $F_p$. The dashed line applies to stellar radius, $R_\star$,
    planetary radius, $R_p$, and the HZ boundaries, $d$. The dotted
    line applies to the planetary density, $\rho_p$.}
  \label{fracplot}
\end{figure}

The various effects described in this section are summarized in the
plot shown in Figure~\ref{fracplot}, which displays the dependencies
of system parameters on the distance ratio,
$d_\star^\prime/d_\star$. The solid line represents the quadratic
relationship of $L_\star$ and $F_p$ on the distance ratio. The dashed
line represents the linear relationship of $R_\star$, $R_p$, and $d$
on the distance ratio. The dotted line represents the cubic
relationship of $\rho_p$ on the distance ratio. Note that, with the
exception of density, the structure of the equations presented here
results in a positive fractional change in parameters if the distance
increases and a negative fractional change if the distance decreases.


\section{Case Studies of Known Habitable Zone Systems}
\label{casestudies}

In this section, we examine three specific systems that are well known
for their HZ planets: TRAPPIST-1, Kepler-186, and
LHS~1140. Specifically, we utilize the revised distances from the {\it
  Gaia} DR2 to re-evaluate the properties of these systems.


\subsection{TRAPPIST-1}
\label{t1}

The TRAPPIST-1 planetary system was first discovered to harbor three
planets by \citet{gil16}, with an additional four later found
\citep{gil17}. {\it K2} observations were used to both confirm and
verify the orbital period of the outer planet \citep{lug17}. The
primary source of interest in the system is due to three of the
planets residing within the HZ of the host star \citep{bol17}. Mass
determinations for the planets were achieved by \citet{gri18} using
Transit Timing Variations (TTVs) leading to greatly improved density
estimates. Furthermore, the density estimates have been used to model
the atmospheres and interiors, leading to conclusions such as that of
\citet{wol17} that planet e has the highest likelihood of liquid
surface water, and the estimates of volatile budgets by
\citet{unt18}. These modeling efforts are further complicated by the
potential for significant atmospheric mass loss due to the relatively
high XUV radiation environment created by the activity of the host
star \citep{rot17,whe17}. The relative closeness of the TRAPPIST-1
system combined with the intrinsic scientific value of the terrestrial
planets make follow-up observations of the system highly likely.

Prior to DR2 becoming available, the properties of the TRAPPIST-1 host
star were known as $d_\star = 12.14 \pm 0.12$~pcs, $T_\mathrm{eff} =
2516 \pm 41$~K, $L_\star = 0.000522 \pm 0.000019$~$L_\odot$, and
$R_\star = 0.121 \pm 0.003$~$R_\odot$ \citep{van18}. DR2 provides a
new parallax for TRAPPIST-1 of $p = 80.4512 \pm 0.1211$~mas, which
corresponds to a revised distance $d_\star = 12.43 \pm
0.02$~pcs. Using the methodology in Section~\ref{distance}, the
luminosity has increased by 5.5\% ($L_\star = 0.000551 \pm
0.000019$~$L_\odot$) and the stellar radius has increased by 2.7\%
($R_\star = 0.124 \pm 0.003$~$R_\odot$).

\begin{deluxetable*}{lccccccc}
  \tablecolumns{8}
  \tablewidth{0pc}
  \tablecaption{\label{fractab} Fractional changes in system properties}
  \tablehead{
    \colhead{System} &
    \colhead{$d_\star^\prime/d_\star$} &
    \colhead{$\Delta L_\star / L_\star$} &
    \colhead{$\Delta F_p / F_p$} &
    \colhead{$\Delta R_\star / R_\star$} &
    \colhead{$\Delta R_p / R_p$} &
    \colhead{$\Delta \rho_p / \rho_p$} &
    \colhead{$\Delta d / d$}
  }
  \startdata
  TRAPPIST-1 & 1.024 & +0.055 & +0.055 & +0.027 & +0.027 & -0.077 & +0.027 \\
  Kepler-186 & 1.176 & +0.382 & +0.382 & +0.176 & +0.176 & -0.385 & +0.176 \\
  LHS~1140   & 1.202 & +0.440 & +0.440 & +0.200 & +0.200 & -0.421 & +0.200
  \enddata
\end{deluxetable*}

\begin{figure*}
  \begin{center}
    \begin{tabular}{cc}
      \includegraphics[angle=270,width=8.5cm]{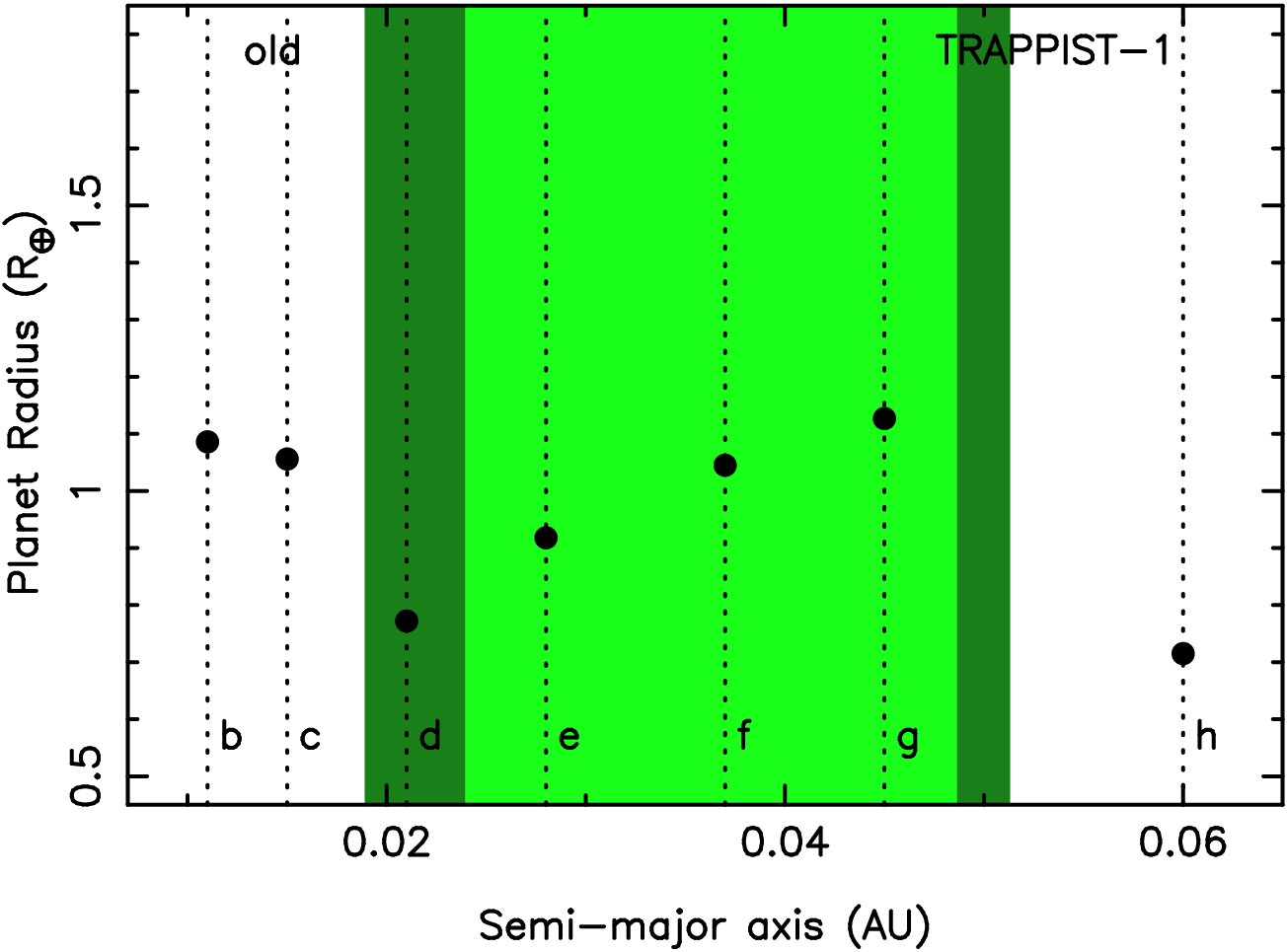} &
      \includegraphics[angle=270,width=8.5cm]{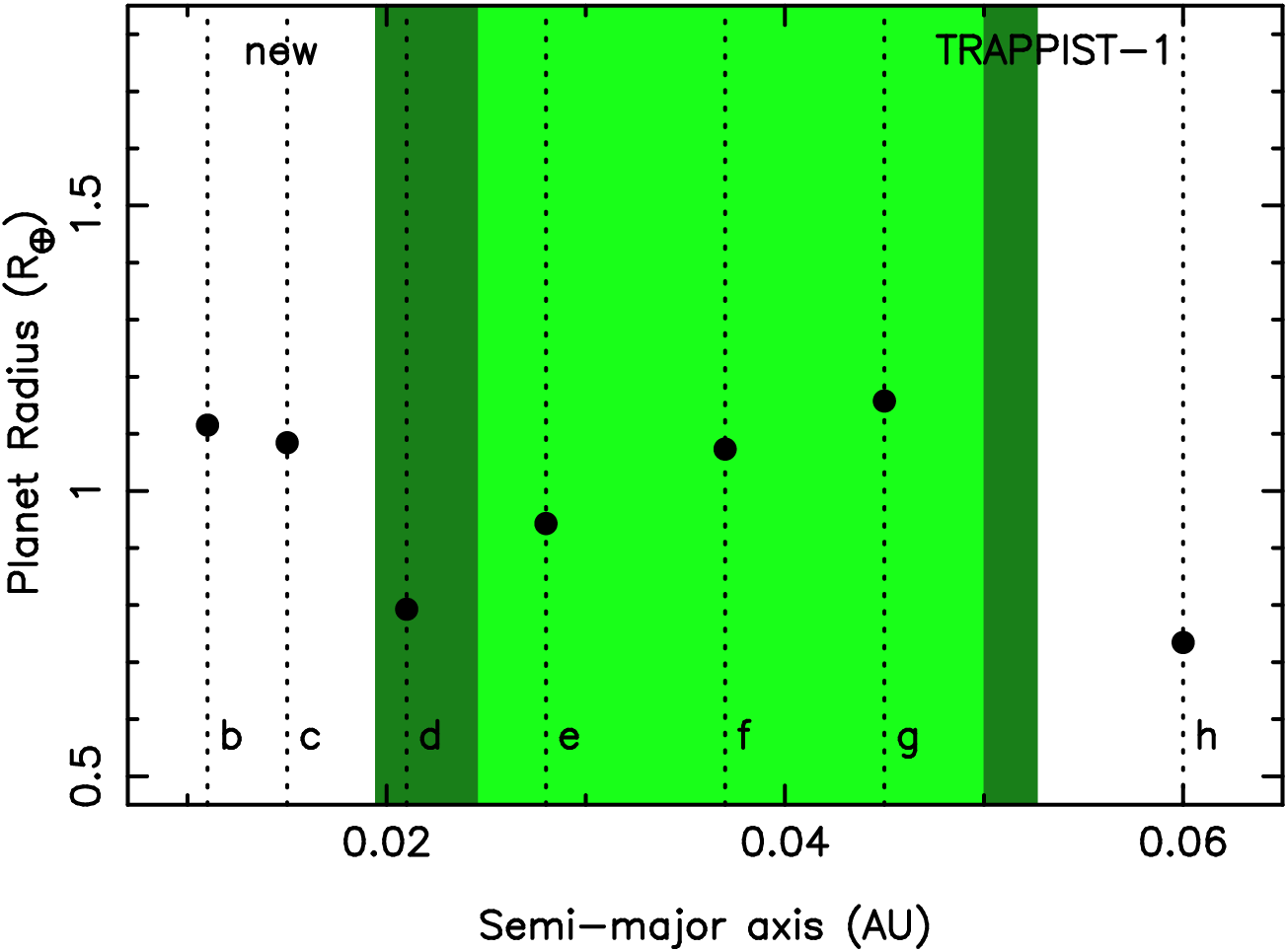} \\
      \includegraphics[angle=270,width=8.5cm]{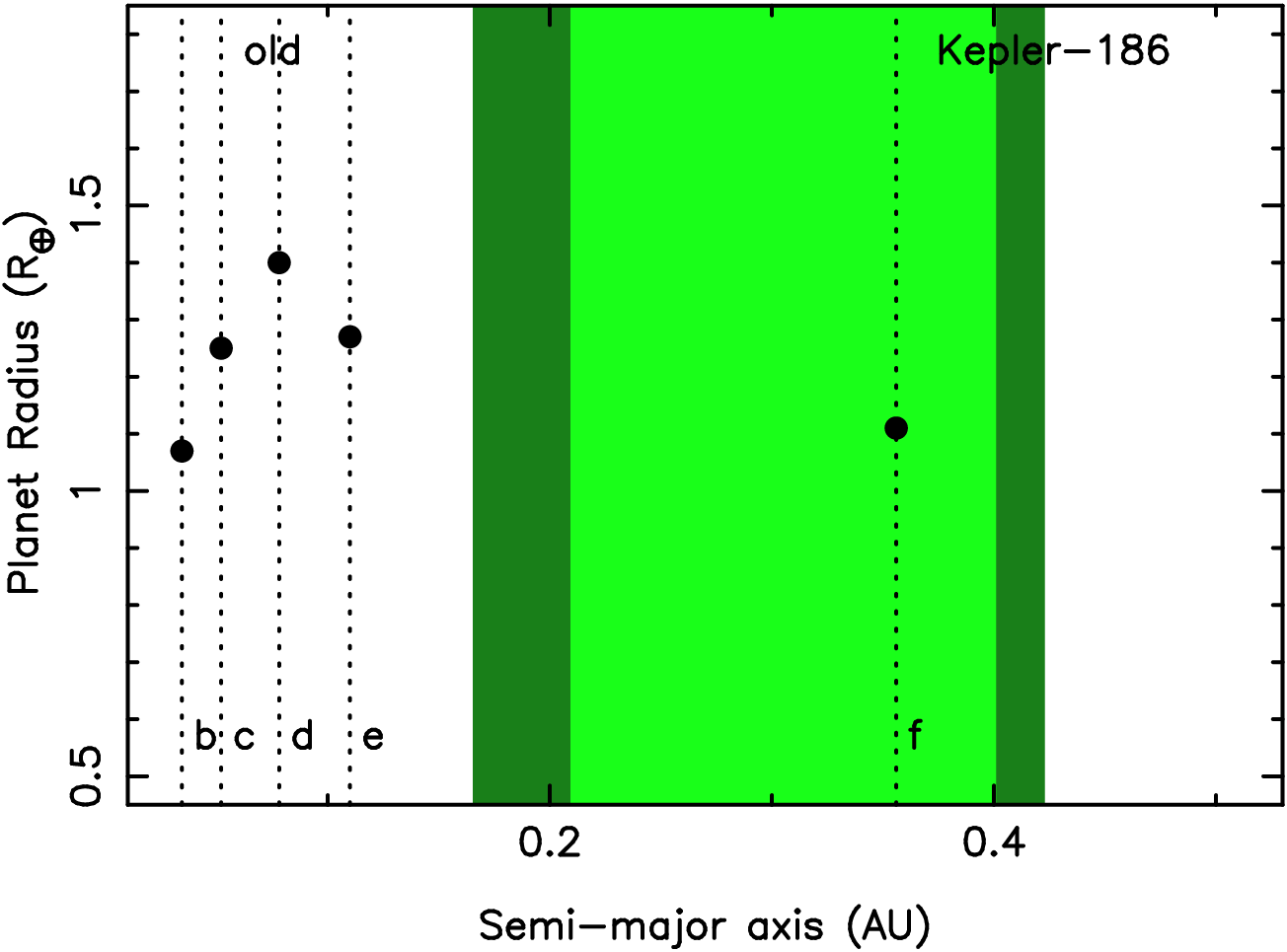} &
      \includegraphics[angle=270,width=8.5cm]{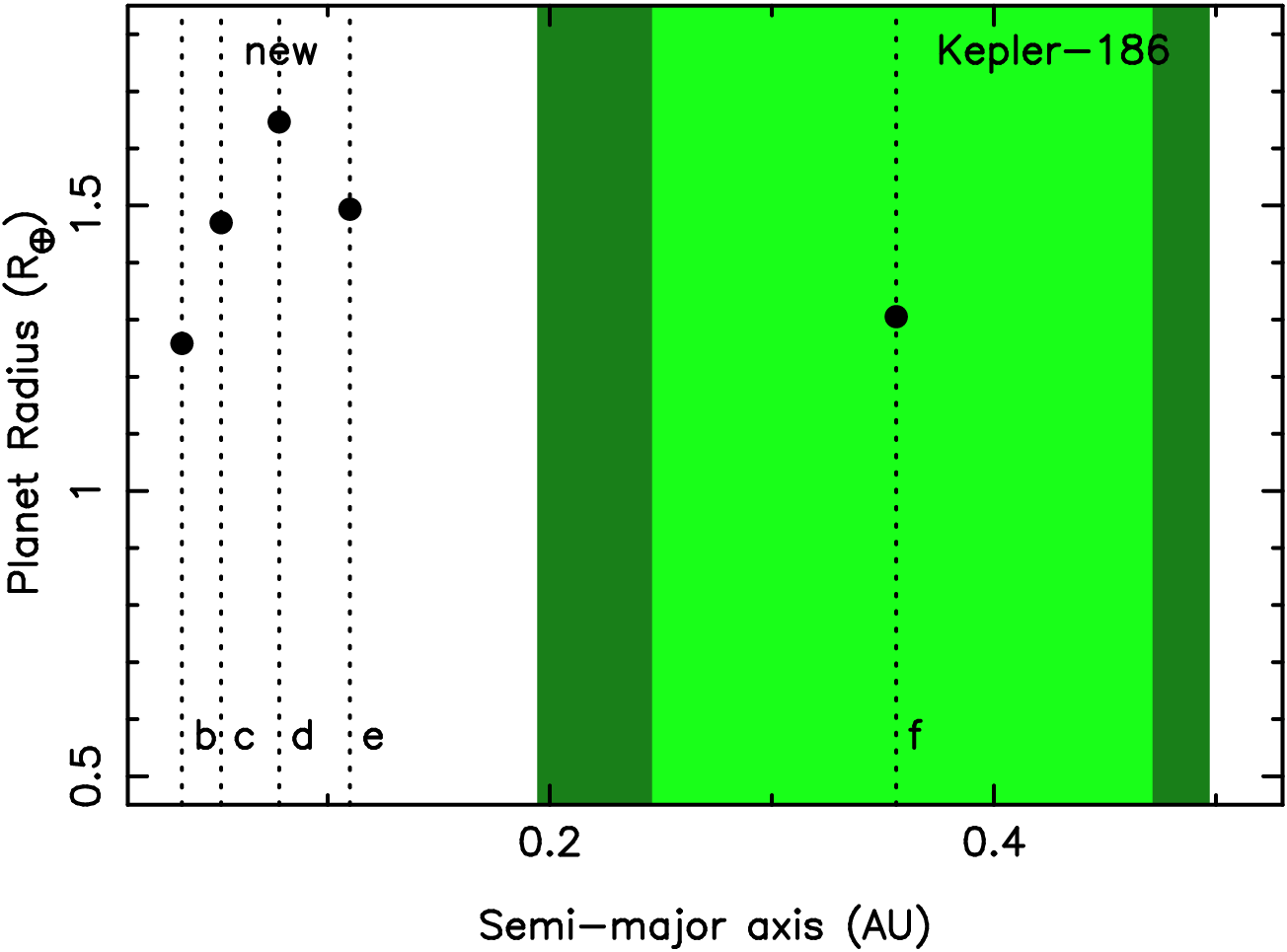} \\
      \includegraphics[angle=270,width=8.5cm]{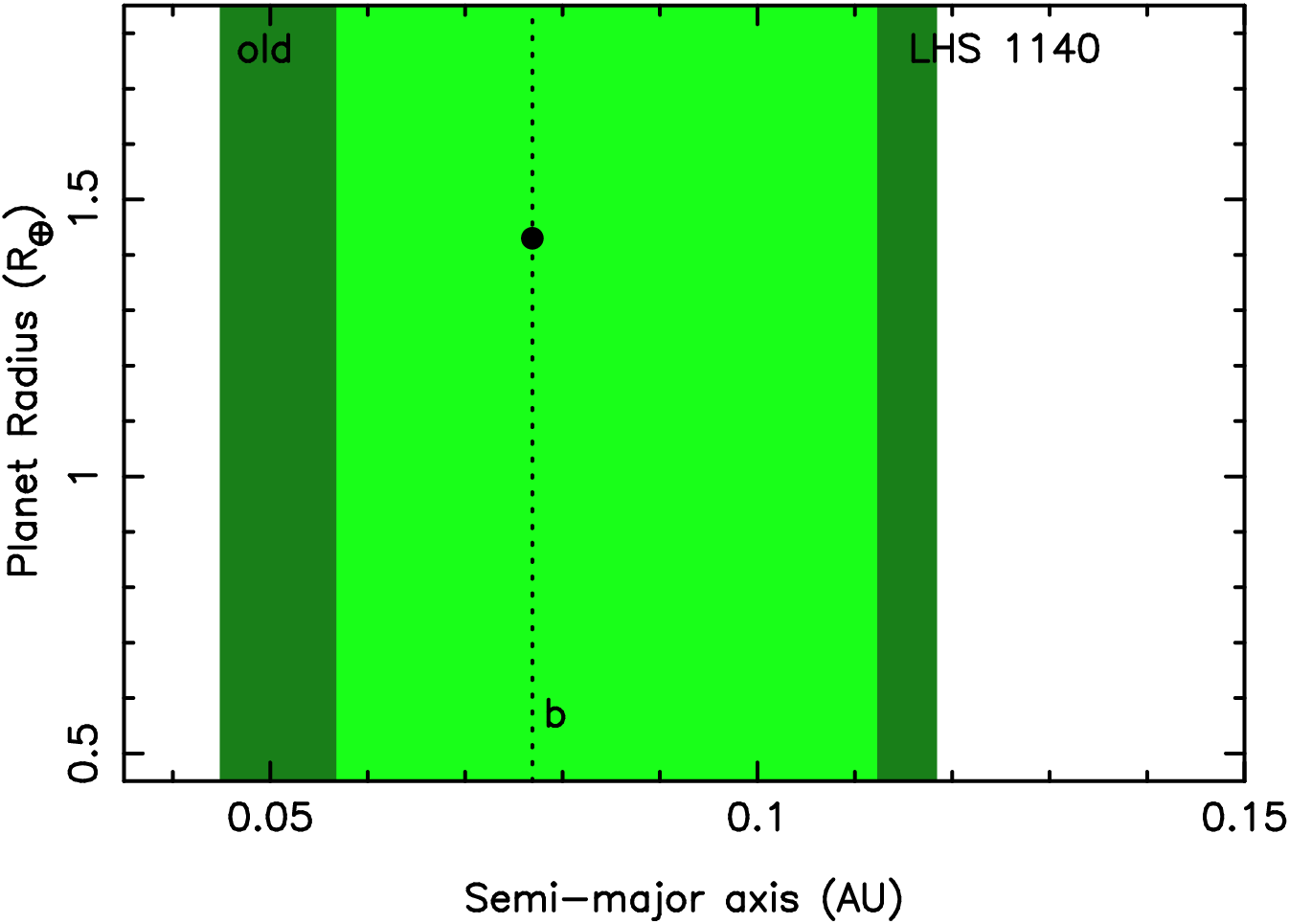} &
      \includegraphics[angle=270,width=8.5cm]{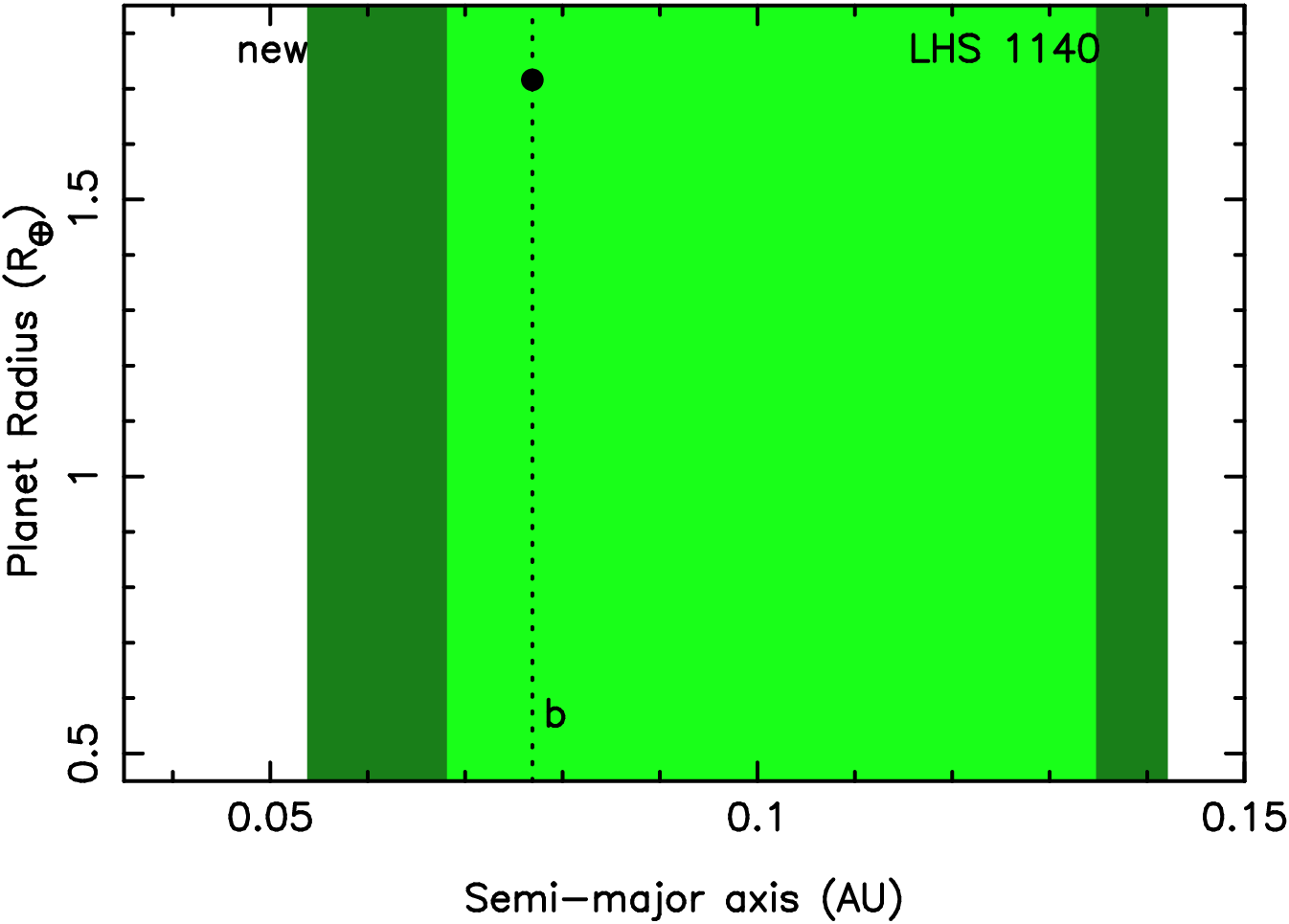}
    \end{tabular}
  \end{center}
  \caption{The changes to the planet sizes and HZ distances resulting
    from the revised DR2 stellar distances for the three case studies
    discussed in Section~\ref{casestudies}. The left column plots
    correspond to the old stellar distances estimates and the right
    column plots correspond to the revised DR2 distances. The top,
    middle, and bottom rows are for the TRAPPIST-1, Kepler-186, and
    LHS~1140 systems respectively. The light-green represents the
    conservative HZ and the dark-green represents the optimistic
    extension to the HZ.}
  \label{updateplots}
\end{figure*}

The consequences of the revised stellar parameters for TRAPPIST-1 are
that the planetary radii increase by 2.7\%, the bulk densities
decrease by 7.7\%. The planetary masses remain unaffected in this case
since they were determined via TTVs. Table~\ref{fractab} summarizes
the fractional change in the system properties for all three case
study systems as a result of the DR2 distance revisions. The changes
are demonstrated graphically in Figure~\ref{updateplots} which shows
the planetary radii and the location of the HZ boundaries (light-green
is the CHZ, dark-green is the OHZ) for the old distance and for the
newly revised distances. The HZ planets of the TRAPPIST-1 system
retain their status as HZ planets when accounting for the new stellar
distance.

The reduction in bulk density of the planets has profound consequences
for their volatile inventory. \citet{unt18} argue that the relatively
low densities of the f and g planets are best explained by formation
beyond the snow line and subsequent migration. They further determine
that a water-rich environment with no continents will lead to a
suppression of biosignatures due to the relative lack of geochemical
cycles involving, for example, carbon and phosphorus. The decrease in
density implies that the volatile content of the TRAPPIST-1 planets is
even higher than previously thought, and indicates that a reanalysis
of the planetary interiors should be considered to properly evaluate
their potential for observations involving biosignatures.


\subsection{Kepler-186}
\label{k186}

The Kepler-186 system was first validated as a confirmed planetary
system due to its multi-planet nature \citep{lis14,row14}, at which
time it was known to contain four planets. The fifth planet was
detected and confirmed by \citet{qui14}, and generated considerable
interest due to its radius of only 1.11~$R_\oplus$ and its location in
the HZ of the host star. At that time, Kepler-186~f was considered the
most ``Earth-like'' in terms of size and insolation flux, leading to
models of the possible energy budgets and habitability potential
\citep{bol14}.

The stellar parameters for Kepler-186 were described by \citet{qui14}
as follows: $d_\star = 151 \pm 18$~pcs, $T_\mathrm{eff} = 3788 \pm
54$~K, $L_\star = 0.0412 \pm 0.0090$~$L_\odot$, and $R_\star = 0.472
\pm 0.052$~$R_\odot$. The DR2 parallax of $p = 5.6020 \pm 0.0243$~mas
equates to a revised distance of $d_\star = 177.51 \pm 0.79$~pcs. The
new distance translates in a luminosity increase of 38.2\% ($L_\star =
0.0569 \pm 0.0090$~$L_\odot$) and a stellar radius increase of 17.6\%
($R_\star = 0.555 \pm 0.052$~$R_\odot$. A refit of the stellar
parameters of Kepler host stars by \citet{ber18} used DR2 distances
and included an isochrone fitting methodology that produced a revised
effective temperature and stellar radius of $T_\mathrm{eff} = 3748 \pm
75$~K and $R_\star = 0.56 \pm 0.02$~$R_\odot$ respectively. Thus, the
stellar radius derived in this work is in close agreement with the
radius derived by \citet{ber18}.

The 17.6\% increase in stellar radius also raises the radius of planet
f to 1.31~$R_\oplus$. This likely means that the planet is still
within the terrestrial regime \citep{chen17}, although it is more
comparable to a super-Earth than a true Earth analog. The mass of the
planet has not been reliably measured due to the lack of detectable
TTV or RV signals, thus the true impact on the bulk density of the
planet is unknown. For a constant mass, the density would decrease by
38.5\%, so this may be considered an upper limit to the density
drop. The HZ boundaries also increase by 17.6\% which is a significant
change in the radiation environment of the system. Fortunately planet
f was previously located at the outer edge of the CHZ and so the
planet has changed from an Earth-size planet at the HZ outer edge to a
super-Earth in the middle of the HZ (see the second row of
Figure~\ref{updateplots}). The modifications to the Kepler-186 system
are summarized in Table~\ref{fractab}.


\subsection{LHS~1140}
\label{l1140}

The planet orbiting LHS~1140 was discovered by \citet{dit17} and
determined to be a rocky super-Earth in the HZ of the host star. The
transit data yielded a radius of 1.43~$R_\oplus$ and the RV data
yielded a mass of 6.65~$M_\oplus$, the combination of which yields a
bulk density of 12.5~g/cm$^3$. As for the TRAPPIST-1 system, the
relative proximity of LHS~1140 and the interest in the planet ensure
that the system will be a valued target for follow-up observations
designed to characterize the atmosphere \citep{che17}.

The stellar parameters provided by \citet{dit17} are $d_\star = 12.47
\pm 0.42$~pcs, $T_\mathrm{eff} = 3131 \pm 100$~K, $L_\star = 0.002981
\pm 0.00021$~$L_\odot$, and $R_\star = 0.186 \pm 0.013$~$R_\odot$. The
DR2 parallax is $p = 66.6996 \pm 0.0674$~mas which translates into a
revised distance of $d_\star = 14.993 \pm 0.016$~pcs. The increase is
stellar distance has the dramatic effect of increasing the luminosity
by 44\% ($L_\star = 0.004293 \pm 0.00021$~$L_\odot$) and the stellar
radius by 20\% ($R_\star = 0.223 \pm 0.013$~$R_\odot$).

The update to the stellar parameters for LHS~1140 results in an
increased planetary radius of 1.72~$R_\oplus$. Normally such a
planetary size would indicate that the planet is not rocky
\citep{rog15} and is more likely to fall into the mini-Neptune
category. Assuming the mass of planet is unchanged, the reduction in
bulk density is 42.1\%, resulting in a revised density of $\rho =
8.8$~g/cm$^3$. The revised density indicates that either the planet is
an exceptionally large super-Earth or that it is a mini-Neptune with a
heavy core. Furthermore, the increase of 20\% in the HZ boundaries
pushes the planet to the near-side of the CHZ. The third row of
Figure~\ref{updateplots} demonstrates the dramatic change to the
properties of the planet. The modifications to the LHS~1140 system are
summarized in Table~\ref{fractab}.


\section{Discussion}
\label{discussion}

The scaling laws presented here for modifying the stellar and
planetary parameters provide a first-order estimate of the expected
changes. However, a more thorough approach will utilize an isochrone
fitting methodology, making use of isochrones such as those compiled
in the ``Dartmouth Stellar Evolution Database'' \citep{dot08}. This is
particularly important because an isochrone re-fitting of the stellar
parameters will likely result in a new value for the stellar
mass. \citet{sta17,sta18} suggest that the masses of host stars can
also be determined through a combination of {\it Gaia} parallaxes and
precision photometry that leads to a measurement of stellar surface
gravity. If a revised mass measurement for a star is found to have
increased from previous measurements, there will be several
consequences for the associated planets, such as increasing the
semi-major axis for a given orbital period, and increasing planetary
masses that are determined using the RV method. Such changes could
help to mitigate decreases to the planetary bulk density and
increases to the incident flux on the planet due to the increased
stellar luminosity.

It should be noted that changes to the stellar luminosity are not the
only factors that affect the habitability of planets within the
HZ. The intrinsic properties of the planet are a critical aspect of
assessing the potential surface conditions, and indeed the extent of
the HZ can be altered by a range of atmospheric circularization
processes and relationships to geophysical cycles
\citep{abb16,haq16,kop17,ram17,haq18,ram18} as well as planetary
rotation rates \citep{yan14,lec15,kop16}. The orbital properties of
the planet also play an important role in the atmospheric dynamics and
surface conditions, such as the orbital eccentricity
\citep{wil02,kan12b}. The scaling laws presented in
Section~\ref{distance} do not directly affect these planetary
properties with the exception of how changes planetary density can
influence surface gravity and atmospheric scale height.

As seen in Section~\ref{casestudies}, the overall effect of changes in
stellar distance for known HZ planets can vary substantially without
an obvious dependence on distance. Kepler-186 has a large correction
for the distance that is not unexpected given that the star is
relatively far away. However, TRAPPIST-1 and LHS~1140 have
dramatically different alterations to their distances and system
properties, despite the fact that they are both relatively close. The
explanation for this is that the TRAPPIST-1 system has been known for
longer and has been studies in much greater detail than the LHS~1140
star, and thus the overall properties of the star, including distance,
have been better characterized.


\section{Conclusions}
\label{conclusions}

Although the dependence of planetary properties on stellar parameters
is well known, the consequences of changing the stellar distance are
less often considered. As demonstrated here, the transition from
poorly to accurately constrained stellar distances can have a serious
impact on planetary properties. This is especially important for those
targets that are expected to form the core sample of expensive
follow-up observations that attempt atmospheric characterization and
those planets that become the subject of detailed climate modeling
simulations.

Perhaps the most important planetary property that changes as a result
of stellar distance is the bulk density. For terrestrial planets in
the HZ, the density together with stellar abundances has become a key
diagnostic in modeling planetary interiors and potential volatile
content \citep{hin18}. For the analysis of the TRAPPIST-1 system
(Section~\ref{t1}) the overall changes to the system are moderate, but
the decrease in the planetary bulk densities is relatively large,
indicating that the planets f and g may indeed be aqua-planets with no
continents, and planet e may also fall into this category. The largest
effect was for LHS~1140b, whose density plummets whilst the received
flux dramatically increases. Such changes to these planets will have a
significant effect on the surface gravity, atmospheric scale height,
and therefore the feasibility of transmission spectroscopy
observations \citep{mor17,bat18}. Thus, as distances are further
improved through {\it Gaia} data releases, the target list to be
studied with precious observing resources will become increasingly
robust.


\section*{Acknowledgements}

The author would like to thank Alma Ceja, Natalie Hinkel, and Eric
Wolf for useful feedback. The author would also like to thank the
anonymous referee, whose comments greatly improved the quality of the
paper This work has made use of data from the European Space Agency
(ESA) mission {\it Gaia}, processed by the {\it Gaia} Data Processing
and Analysis Consortium (DPAC). Funding for the DPAC has been provided
by national institutions, in particular, the institutions
participating in the {\it Gaia} Multilateral Agreement. This research
has also made use of the Habitable Zone Gallery at
\url{hzgallery.org}, and the NASA Exoplanet Archive, which is operated
by the California Institute of Technology, under contract with the
National Aeronautics and Space Administration under the Exoplanet
Exploration Program. The results reported herein benefited from
collaborations and/or information exchange within NASA's Nexus for
Exoplanet System Science (NExSS) research coordination network
sponsored by NASA's Science Mission Directorate.


\end{document}